\documentclass[%
 reprint,
superscriptaddress,
 amsmath,amssymb,
aps,
prx,
]{revtex4-2}

\usepackage{graphicx}
\usepackage{dcolumn}
\usepackage{bm}
\usepackage{siunitx}
\usepackage{textcomp}

\graphicspath{{physrev_figures/}}

\AtBeginDocument{\usepackage{booktabs}}

\begin{document}


\title{Extracting Interpretable Physical Parameters from Spatiotemporal Systems using Unsupervised Learning}

\author{Peter Y. Lu}
\email{lup@mit.edu}
\affiliation{%
 Department of Physics, Massachusetts Institute of Technology, Cambridge, MA 02139, USA
}%
\author{Samuel Kim}%
\affiliation{%
 Department of Electrical Engineering and Computer Science, Massachusetts Institute of Technology, Cambridge, MA 02139, USA
}%
\author{Marin Solja\v{c}i\'c}%
\affiliation{%
 Department of Physics, Massachusetts Institute of Technology, Cambridge, MA 02139, USA
}%

\date{\today}

\begin{abstract}
    Experimental data is often affected by uncontrolled variables that make analysis and interpretation difficult. For spatiotemporal systems, this problem is further exacerbated by their intricate dynamics. Modern machine learning methods are particularly well-suited for analyzing and modeling complex datasets, but to be effective in science, the result needs to be interpretable. We demonstrate an unsupervised learning technique for extracting interpretable physical parameters from noisy spatiotemporal data and for building a transferable model of the system. In particular, we implement a physics-informed architecture based on variational autoencoders that is designed for analyzing systems governed by partial differential equations (PDEs). The architecture is trained end-to-end and extracts latent parameters that parameterize the dynamics of a learned predictive model for the system. To test our method, we train our model on simulated data from a variety of PDEs with varying dynamical parameters that act as uncontrolled variables. Numerical experiments show that our method can accurately identify relevant parameters and extract them from raw and even noisy spatiotemporal data (tested with roughly 10\% added noise). These extracted parameters correlate well (linearly with $R^2 > 0.95$) with the ground truth physical parameters used to generate the datasets. We then apply this method to nonlinear fiber propagation data, generated by an ab-initio simulation, to demonstrate its capabilities on a more realistic dataset. Our method for discovering interpretable latent parameters in spatiotemporal systems will allow us to better analyze and understand real-world phenomena and datasets, which often have unknown and uncontrolled variables that alter the system dynamics and cause varying behaviors that are difficult to disentangle.

\end{abstract}

\maketitle


\section{\label{sec:intro}Introduction}

Physics has traditionally relied upon human ingenuity to identify key variables, discover physical laws, and model dynamical systems. With the recent explosion of available data coupled with advances in machine learning, fundamentally new methods of discovery are now possible. However, a major issue with applying these novel techniques to scientific and industrial applications is their interpretability: neural networks and deep learning are often seen as inherently black box methods. To make progress, we must incorporate scientific domain knowledge into our network architecture design and algorithms without sacrificing the flexibility provided by deep learning models \cite{Karpatne:ec,1710.11431,1901.11103}. In this work, we show that we can leverage unsupervised learning techniques in a physics-informed architecture to build models that learn to both identify relevant interpretable parameters and perform prediction. Because relevant parameters are necessary for predictive success, the two tasks of extracting parameters and creating a predictive model are closely linked, and we exploit this relationship to do both using a single architecture.

We focus our attention on spatiotemporal systems with dynamics governed by partial differential equations (PDEs). These systems are ubiquitous in nature and include physical phenomena in fluid dynamics and electromagnetism. Recently, there has been significant interest in the data-driven analysis and discovery of PDEs: e.g.\ explicitly identifying PDEs using sparse linear regression with a library of possible terms \cite{Rudy:2017cs}, using a convolutional architecture with symbolic regression to identify PDEs \cite{pmlr-v80-long18a,LONG2019108925}, and representing PDE solutions as neural networks to solve and identify PDEs \cite{RAISSI2019686,BERG2019239,JMLR:v19:18-046}. However, previous works on PDE discovery and parameter extraction often assume the entire dataset is governed by the same dynamics and also explicitly provide the key dynamical parameters (with potentially unknown values). In more complex scenarios, we may have limited control over the systems that we are studying and yet still want to model them and extract relevant physical features from their dynamics. If we attempt to study such systems by na\"ively training a predictive model, we are likely to fail in one of two ways: first, a single explicit PDE model will be unable to capture the variations in the dynamics caused by uncontrolled variables in the data, and second, a generic deep learning method for time-series prediction such as long short-term memory (LSTM)-based models \cite{hochreiter1997long,doi:10.1162/089976600300015015} will not be interpretable or provide any physical insight and may also result in unphysical solutions at later times due to overfitting. To avoid these problems and gain a better understanding of the physical system, we must first identify important parameters or variables that are uncontrolled and that change in the raw data, producing varying dynamics. Recent work on learning parametric PDEs has taken steps toward addressing this issue \cite{doi:10.1137/18M1191944}. We will use an unsupervised learning method to automate the process of determining the relevant parameters that control the system dynamics and constructing a predictive model---all without requiring information on the form of the governing PDE.

We propose a model architecture (Fig.\ \ref{fig:arch}(a)) based on variational autoencoders (VAEs) \cite{1312.6114}. VAEs are widely used for dimensionality reduction and unsupervised learning tasks \cite{Goodfellow-et-al-2016} and have been shown to be effective for studying a wide variety of physical phenomena, e.g.\ discovering simple representations of systems in classical and quantum mechanics \cite{1807.10300}, modeling protein folding and molecular dynamics \cite{PhysRevE.97.062412,Wehmeyer:2018hy}, and identifying condensed matter phase transitions \cite{PhysRevE.96.022140,1705.09524}. In terms of interpretability, the VAE architecture and its derivatives have also been shown to disentangle independent factors of variation \cite{higgins2017beta,1804.03599,NIPS2018_7527}. The choice of a VAE-based architecture is motivated both by their prior success in extracting useful representations and by their strong theoretical foundation (Appendix \ref{apx:vaereg}). Prior works have also applied other methods of parameter identification, such as using principal component analysis (PCA) as a post-processing step with a standard autoencoder \cite{1807.09244}. This, however, relies heavily on the poorly understood implicit regularization of the neural network architecture rather than the explicit regularization of a VAE, and, in our experience, VAEs produce more consistent and interpretable results.

Our architecture consists of an encoder (Fig.\ \ref{fig:arch}(b)) that extracts physical parameters characterizing the system dynamics and a decoder (Fig.\ \ref{fig:arch}(c)) that acts as a predictive model and propagates an initial condition forward in time given the extracted parameters. This differs from a traditional VAE due to the additional initial condition provided to the decoder, allowing the encoder to focus on extracting latent parameters that parameterize the dynamics of the system rather than the physical state. Our architecture can be thought of as a conditional VAE \cite{NIPS2015_5775}, although only the decoder is conditional. While similar architectures have been recently proposed for physical systems such as interacting particles \cite{1807.09244} and moving objects \cite{xue2018visual}, our model is specifically designed to study spatiotemporal phenomena, which have a continuous set of degrees of freedom.

To take advantage of the spatiotemporal structure of PDE-governed systems, we use convolutional layers---commonly employed in image recognition tasks \cite{NIPS2012_4824,He_2016_CVPR} to efficiently represent local features---in both the encoder and decoder portions of our architecture. The translation invariance of the convolutions allows us to train on small patches of data and then evaluate on larger systems with arbitrary boundary conditions. In the decoder, the convolutional layers are placed in a recurrent architecture to represent time propagation---analogous to a PDE solver with a finite difference approximation \cite{pmlr-v80-long18a}. In addition, our architecture efficiently parameterizes PDE propagation by dynamically generating the convolutional kernels and biases of the decoder using the extracted latent parameters from the encoder. In this way, the latent parameters directly control the local propagation of the physical states in the decoder, resulting in more stable model predictions and a more physical encoding of the dynamics. These architecture choices provide the key physics-informed inductive biases that enhance the interpretability of the extracted parameters and ensure a physically reasonable predictive model.

To demonstrate the capabilities of this approach, we test our method on simulated data from PDE models for chaotic wave dynamics, optical nonlinearities, and convection and diffusion (Sec.\ \ref{sec:data}). These numerical experiments show that our method can accurately identify and extract relevant physical parameters that characterize variations in the observed dynamics of a spatiotemporal system (Sec.\ \ref{sec:exp_extr}), while at the same time construct a flexible and transferable predictive model (Sec.\ \ref{sec:exp_pred}). We further show that the parameter extraction is robust to noisy data and can still be effective for chaotic systems where accurate prediction is difficult. Finally, we apply this method to nonlinear optical fiber propagation using data generated from an ab-initio electromagnetic simulation to test the model on a more realistic dataset (Sec.\ \ref{sec:fiber}). The goal of our approach is to provide an additional tool for studying complex spatiotemporal systems when there are unknown and uncontrolled variables present.

\section{\label{sec:arch}Model Architecture}

\begin{figure*}
    \vspace{1em}
    \centering
    \includegraphics[width=\linewidth]{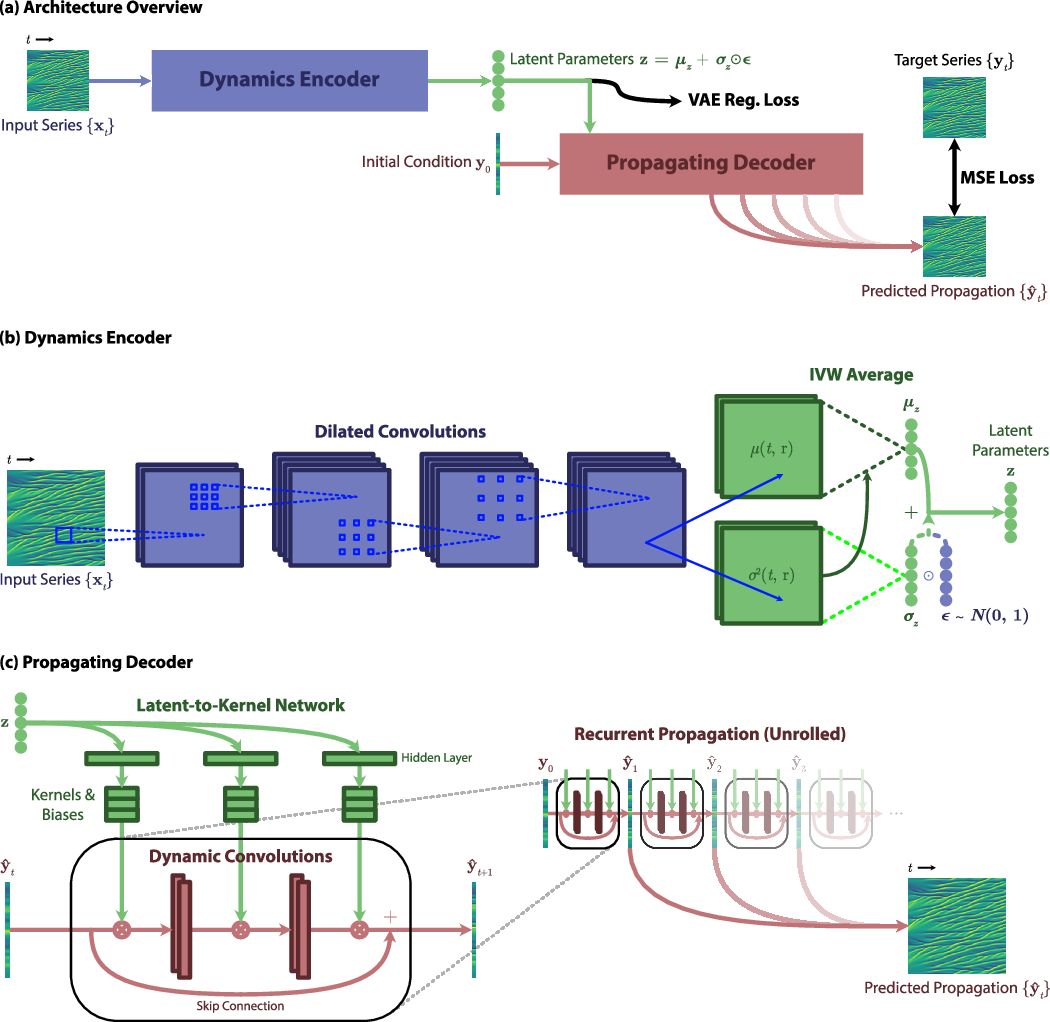}
    \caption{VAE-based model architecture. (a) The architecture consists of the dynamics encoder (DE) and the propagating decoder (PD) with kernels and biases given by a latent-to-kernel network. (b) The DE extracts the latent distribution parameters $\mu_z$ and $\sigma_z$ from the input series $\{\mathbf{x}_t\}_{t=0}^{T_x}$ using dilated convolutions and inverse-variance weighted (IVW) averaging. During training, the latent parameters are sampled from the extracted distribution $z \sim \mathcal N(\mu_z, \sigma_z^2)$. (c) The PD then uses a fully-connected latent-to-kernel network to map the latent parameters $z$ to the kernels and biases of the dynamic convolutional layers, which are used in a recurrent fashion to predict the propagation of the system $\{\mathbf{\hat y}_t\}_{t=1}^{T_y}$ starting from the initial condition $\mathbf{\hat y}_0 = \mathbf{y}_0$. The model is trained end-to-end using the mean squared error (MSE) loss between the predicted propagation series $\{\mathbf{\hat y}_t\}_{t=1}^{T_y}$ and target series $\{\mathbf{y}_t\}_{t=1}^{T_y}$ along with VAE regularization. Time-series limits are dropped in the figure labels for conciseness.}
    \label{fig:arch}
\end{figure*}

Our model (Fig.\ \ref{fig:arch}) has an encoder--decoder architecture based on a variational autoencoder (VAE) \cite{1312.6114,higgins2017beta}. Given a dataset of time-series from a spatiotemporal system, the {\em dynamics encoder} (DE) extracts latent parameters which parameterize the varying dynamics in the dataset. These latent parameters are then used by the {\em propagating decoder} (PD) to simulate the system given an initial condition and boundary conditions. During training, the model is optimized to match the output of the PD to a time-series example from the dataset. The goal of the VAE architecture is to allow the PD to push the DE to extract useful and informative latent parameters.

For training, the network requires time-series data that are grouped in pairs: the {\em input series} $\{\mathbf{x}_t\}_{t=0}^{T_x}$ is the input to the DE, and the {\em target series} $\{\mathbf{y}_t\}_{t=0}^{T_y}$ provides the initial condition $\mathbf{y}_0$ and the training targets $\{\mathbf{y}_t\}_{t=1}^{T_y}$ for the PD. Each pair of time-series $(\{\mathbf{x}_t\}_{t=0}^{T_x},\{\mathbf{y}_t\}_{t=0}^{T_y})$ must follow the same dynamics and thus correspond to the same latent parameters. We can construct such a dataset from the raw data by cropping each original time-series to produce a pair of smaller time-series. This cropping can be performed randomly in both the time and space dimensions, which allows the network to train on a reduced system size while still making use of all the available data. In our examples, we also choose to crop dynamically during training, akin to data augmentation methods used in image recognition \cite{He_2016_CVPR}.

In detail, the DE network takes the full input series $\{\mathbf{x}_t\}_{t=0}^{T_x}$ and outputs a mean $\mu_{z|\mathbf{x}}$ and a variance $\sigma^{2}_{z|\mathbf{x}}$, which we will henceforth refer to as $\mu_z$ and $\sigma^{2}_z$ for compactness, representing a normal distribution for each latent parameter $z$. During training, each $z$ is sampled from its corresponding distribution $\mathcal N(\mu_z, \sigma^{2}_z)$ using the VAE reparameterization trick: $z = \mu_z + \sigma_z\epsilon$, where $\epsilon \sim \mathcal N(0,1)$ is independently sampled for every training example during each training step. During evaluation, we simply take $z = \mu_z$. These parameters $z$ along with an initial condition---the first state $\mathbf{y}_0$ in the target series---are then used by the PD network to predict the future propagation of the system. The predicted propagation series $\{\mathbf{\hat y}_t\}_{t=1}^{T}$ produced by the PD can be computed up to an arbitrary future time $T$, where $T=T_y$ during training to match the target series. By providing the PD with an initial condition, we allow the DE to focus on encoding parameters that characterize the dynamics of the data rather than encoding a particular state of the system. This is further reinforced by training on randomly cropped pairs of time-series as well as by the VAE regularization term (Appendix \ref{apx:vaereg}).

The full architecture is trained end-to-end using a mean-squared error loss between the predicted propagation series $\{\mathbf{\hat y}_t\}_{t=1}^{T_y}$ from the PD and the target series $\{\mathbf{y}_t\}_{t=1}^{T_y}$. We also add the VAE regularization loss---the KL divergence term $D_\mathrm{KL}(\mathcal N(\mu_z, \sigma^2_z)\;\|\;\mathcal N(0,1))$---which encourages each latent parameter distribution $\mathcal N(\mu_z, \sigma^2_z)$ generated by the DE to approach the standard normal prior distribution $\mathcal N(0,1)$. The total loss function is given by
\begin{align}
	\mathcal L =\, &\frac{1}{T_y}\sum_{t=1}^{T_y} \left(\mathbf{y}_t - \mathbf{\hat y}_t\right)^2\\
	        &+ \beta\,\sum_zD_\mathrm{KL}(\mathcal N(\mu_z, \sigma^2_z)\;\|\;\mathcal N(0,1)),
\end{align}
where $T_y$ is the length of the target series without the initial $\mathbf{y}_0$, and $\beta$ is a regularization hyperparameter which we tune for each dataset. The $\beta$ parameter is key to learning disentangled representations \cite{higgins2017beta,1804.03599}. By using the VAE sampling method and regularizer, we compel the model to learn independent and interpretable latent parameters (Appendix \ref{apx:vaereg}). For additional training details, see Appendix \ref{apx:train}.

The source code for our implementation is available at \url{https://github.com/peterparity/PDE-VAE-pytorch}.

\subsection{Dynamics Encoder (DE)\label{sec:arch/de}}

The dynamics encoder (DE) network is designed to take advantage of existing symmetry or structure in the time-series data. We implement the DE as a deep convolutional network in both the time and space dimensions to allow the network to efficiently extract relevant features. To ensure the DE can handle arbitrary system sizes and time-series lengths, the architecture only contains convolutional layers with a weighted average applied at the output to obtain the latent parameters. The weights for the final averaging are also learned by the network and interpreted as variances so that the overall variance can also be computed. In this way, the network is able to focus on areas of the input series that are most important for estimating the latent parameters, akin to a visual attention mechanism \cite{pmlr-v37-xuc15}.

Explicitly, we first compute local quantities
\begin{align}
    \mu(t,\mathbf{r}) &= f_{\mathrm{DE},\mu}(\{\mathbf{x}_{t'}\}_{t'=0}^{T_x})\\
    \log \sigma^2(t,\mathbf{r}) &= f_{\mathrm{DE},\sigma^2}(\{\mathbf{x}_{t'}\}_{t'=0}^{T_x}),
\end{align}
where $f_{\mathrm{DE},\mu}$ and $f_{\mathrm{DE},\sigma^2}$ are multilayer convolutional networks in space and time (see Appendix \ref{apx:model} for details). Then, instead of using a fully-connected layer to compute the final mean $\mu_z$ and variance $\sigma^2_z$ for each latent parameter, we combine the local quantities by performing an inverse-variance weighted average using weights given by
\begin{equation}
    w(t,\mathbf{r}) = \sigma^{-2}(t,\mathbf{r})/\sum_{t,\mathbf{r}} \sigma^{-2}(t,\mathbf{r})
\end{equation}
to obtain
\begin{align}
    \mu_z &= \sum_{t,\mathbf{r}} w(t,\mathbf{r})\,\mu(t,\mathbf{r})\\
    \sigma^2_z &= C^d/\sum_{t,\mathbf{r}} \sigma^{-2}(t,\mathbf{r}),
\end{align}
where $C$ is a constant chosen to correct for the correlations between nearby points and $d$ is the total number of time and space dimensions of the input. This averaging serves two purposes: it allows the DE to scale to arbitrary system sizes and geometries, and it improves the parameter extraction by placing greater emphasis on regions of high confidence. Assuming that non-overlapping patches should be treated as independent while overlapping patches are increasingly correlated, we take $C=31$ to be the linear size of the receptive field of the convolutional networks $f_{\mathrm{DE},\mu}$ and $f_{\mathrm{DE},\sigma^2}$.

\subsection{Propagating Decoder (PD)\label{sec:arch/pd}}

The propagating decoder (PD) network is designed as a predictive model for spatiotemporal systems. We structure the PD as a multilayered convolutional network $f_\mathrm{PD}$ (see Appendix \ref{apx:model} for details) with a residual skip connection that maps a state $\mathbf{\hat y}_t$ to the next state in the time-series $\mathbf{\hat y}_{t+1}$. Thus, each propagation step is given by
\begin{equation}
    \mathbf{\hat y}_{t+1} = \mathbf{\hat y}_t + f_\mathrm{PD}(\mathbf{\hat y}_t).
\end{equation}
To generate the predicted propagation series $\{\mathbf{\hat y}_t\}_{t=1}^{T_y}$ for comparison with the target series $\{\mathbf{y}_t\}_{t=1}^{T_y}$, we begin with the initial condition $\mathbf{\hat y}_0 = \mathbf{y}_0$ and then recursively apply the PD to propagate $\mathbf{\hat y}_t \to \mathbf{\hat y}_{t+1}$, forming a recurrent network. The PD acts as a physics simulator or, in this case, a PDE integrator with explicit time stepping. This architecture reflects both the spatial and temporal structure of PDE-governed systems and incorporates boundary conditions by properly padding $\mathbf{\hat y}_t$ at each time step before applying the convolutional layers. For example, to use periodic boundary conditions during evaluation, we apply periodic padding at each time step. During training, we treat the edges of the target series---cropped from a full training example---as a boundary condition by using the spatial boundary of each $\mathbf{y}_t$ in the target series to pad the corresponding state $\mathbf{\hat y}_t$ before propagation.

Unlike the convolutional layers of the DE, the kernels and biases for $f_\mathrm{PD}$ are not directly trained. Instead, the kernel weights and biases are a function of the latent parameters $z$. This type of layer is known as a dynamic convolution \cite{NIPS2016_6578} or a cross-convolution \cite{xue2018visual}. Each convolutional kernel and corresponding bias is constructed by a separate fully-connected {\em latent-to-kernel} network that maps the latent parameters to each kernel or bias, forming a multiplicative connection in the PD. Thus, we can interpret the PD convolutional kernels and biases as encoding the dynamics of the system parameterized by $z$.

\section{\label{sec:data}Simulated PDE Datasets}

To study the ability of our architecture to perform parameter extraction, we generate simulated datasets of spatiotemporal systems that have spatially uniform, time-independent local dynamics in a box with periodic boundary conditions, i.e.\ we consider PDEs of the form
\begin{equation}
    \frac{\partial\mathbf{u}(t,\mathbf{r})}{\partial t} = F(\mathbf{u}, \nabla\mathbf{u}, \mathbf{u}^2, (\mathbf{u}\cdot\nabla)\mathbf{u}, \ldots),
\end{equation}
where $F$ is a general space- and time-independent, nonlinear local operator acting on $\mathbf{u}$. This allows us to design an optimized, physics-informed model architecture. We test our model on a variety of spatiotemporal systems by creating the following three datasets that cover linear, nonlinear, and chaotic dynamics as well as giving both 1D and 2D examples. For details on the generation of the simulated datasets, see Appendix \ref{apx:datasets}.

\subsection{1D Kuramoto--Sivashinsky}
The Kuramoto--Sivashinsky equation
\begin{equation}
    \frac{\partial u}{\partial t} = - \gamma\,\partial_x^4 u -\partial_x^2 u -u\,\partial_xu 
\end{equation}
is nonlinear scalar wave equation with a viscosity damping parameter $\gamma$. This is a key example of a chaotic PDE \cite{manneville1985liapounov} due to the instability caused by the negative second derivative term and was originally derived to model laminar flame fronts \cite{kuramoto1978diffusion,sivashinsky1980flame}. The 1D Kuramoto--Sivashinsky dataset has a training set with 5,000 examples and a test set with 10,000 examples.

\subsection{1D Nonlinear Schr\"odinger}
The nonlinear Schr\"odinger equation
\begin{equation}
    i\frac{\partial \psi}{\partial t} =  -\frac12\partial_x^2\psi +\kappa \left|\psi\right|^2\psi
\end{equation}
is a complex scalar wave equation with a cubic nonlinearity controlled by the coefficient $\kappa$. In our data, we represent $\psi = u_1 + iu_2$ as a real two-component vector $\mathbf{u} = (u_1,u_2)$. This equation can be used to model the evolution of wave-packets in nonlinear optics and is known to exhibit soliton solutions \cite{ablowitz_2011}. The 1D nonlinear Schr\"odinger dataset has a training set with 5,000 examples and a test set with 10,000 examples.

\subsection{2D Convection--Diffusion}
The 2D convection--diffusion equation
\begin{equation}
    \frac{\partial u}{\partial t} = D\nabla^2u - \mathbf{v} \cdot \nabla u
\end{equation}
is a linear scalar wave equation consisting of a diffusion term with constant $D$ and a velocity-dependent convection term with velocity field $\mathbf{v}$. The equation describes a diffusing quantity that is also affected by the flow or drift of the system, e.g.\ dye diffusing in a moving fluid. We consider the case of a constant velocity field. The 2D convection--diffusion dataset has a training set with 1,000 examples and a test set with 1,000 examples.

\section{\label{sec:exp}Numerical Experiments}

We perform numerical experiments by training the model on both the original noiseless datasets and the datasets with added $\sigma=0.1$ Gaussian noise---corresponding to 10\% noise relative to the initial conditions. Then, we evaluate the trained models on the full size noiseless test set examples (no cropping). By also training on noisy datasets, we test the robustness of our method and show the effect of noise on the extracted parameters and prediction performance.

\subsection{\label{sec:exp_extr}Parameter Extraction}

\begin{table}
\caption{$R^2$ correlation coefficients from linear fits of the relevant latent parameters (Fig.\ \ref{fig:relparams}) with the ground truth physical parameters for each dataset---both with and without added noise. For the three-parameter convection--diffusion dataset, the diffusion constant $D$ is fit with a corresponding extracted latent parameter, while the drift velocity components $v_x$, $v_y$ are fit with a corresponding two-dimensional subspace of the latent parameters due to the inherent rotational symmetry.}
\begin{ruledtabular}
\begin{tabular}{cccc}
    Dataset                 & Param. & No Noise  & $\sigma=0.1$ Noise\\
    \midrule
    Kuramoto--Sivashinsky   & $\gamma$  & 0.993             & 0.995\\
    Nonlinear Schr\"odinger & $\kappa$  & 0.997             & 0.998\\
    Convection--Diffusion   & $D$       & 0.963             & 0.959\\
    Convection--Diffusion   & $v_x$     & 0.997             & 0.994\\
    Convection--Diffusion   & $v_y$     & 0.998             & 0.996\\
\end{tabular}
\end{ruledtabular}
\label{tab:corr}
\end{table}

\begin{table}
\centering
\caption{$R^2$ correlation coefficients from individual linear fits of the 2D convection--diffusion dataset parameters with each relevant latent parameter (LP). High correlations are bolded, emphasizing the interpretability of the learned latent parameters as either corresponding to the diffusion constant $D$ or the drift velocity components $v_x$, $v_y$. The drift velocity is matched with two latent parameters that form a two-dimensional latent subspace corresponding to the velocity vector.}
\begin{ruledtabular}
\begin{tabular}{c ccc c ccc}
    & \multicolumn{3}{c}{No Noise} && \multicolumn{3}{c}{$\sigma = 0.1$ Noise}\\
    \cmidrule{2-4}\cmidrule{6-8}
        Param.  & LP 1           & LP 2           & LP 5           && LP 1            & LP 2            & LP 3\\
    \midrule
        $D$     & \textbf{0.963} & 0.000          & 0.003          &&  0.003          &  \textbf{0.959} & 0.001\\
        $v_x$   & 0.000          & \textbf{0.205} & \textbf{0.766} &&  \textbf{0.395} &  0.006          & \textbf{0.554}\\
        $v_y$   & 0.001          & \textbf{0.818} & \textbf{0.205} &&  \textbf{0.568} &  0.000          & \textbf{0.473}\\
\end{tabular}
\end{ruledtabular}
\label{tab:CD}
\end{table}

\begin{figure}
    \centering
    \includegraphics[width=\linewidth]{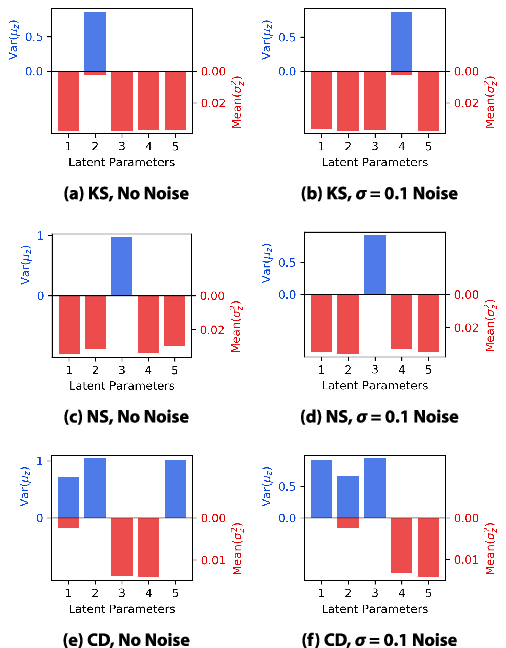}
    \caption{Identification of relevant latent parameters: the variance of $\mu_z$ (blue) and mean of $\sigma_z^2$ (red) for the five latent parameters in the models trained on the (a), (b) 1D Kuramoto--Sivashinsky (KS), (c), (d) 1D nonlinear Schr\"odinger (NS), and (e), (f) 2D convection--diffusion (CD) datasets both with and without added noise. In each case, the model has correctly identified the number of relevant parameters (one for the Kuramoto--Sivashinsky and nonlinear Schr\"odinger datasets, and three for the convection--diffusion dataset), which are characterized by high variance in $\mu_z$ and a low mean $\sigma_z^2$. These relevant latent parameters correspond to interpretable physical parameters that parameterize the dynamics of the system. The other latent parameters with near zero variance in $\mu_z$ and high mean $\sigma_z^2$ have collapsed to the prior and are non-informative. Note that while one would expect these collapsed parameters to have $\sigma_z^2=1$, the actual extracted $\sigma_z^2$ for the collapsed non-informative parameters is less than one. This is an artifact of evaluating the model on a larger system size and longer time-series than the cropped patches used during training (see Appendix \ref{apx:rawresults} for details).}
    \label{fig:relparams}
\end{figure}

\begin{figure*}
    \centering
    \includegraphics[width=\linewidth]{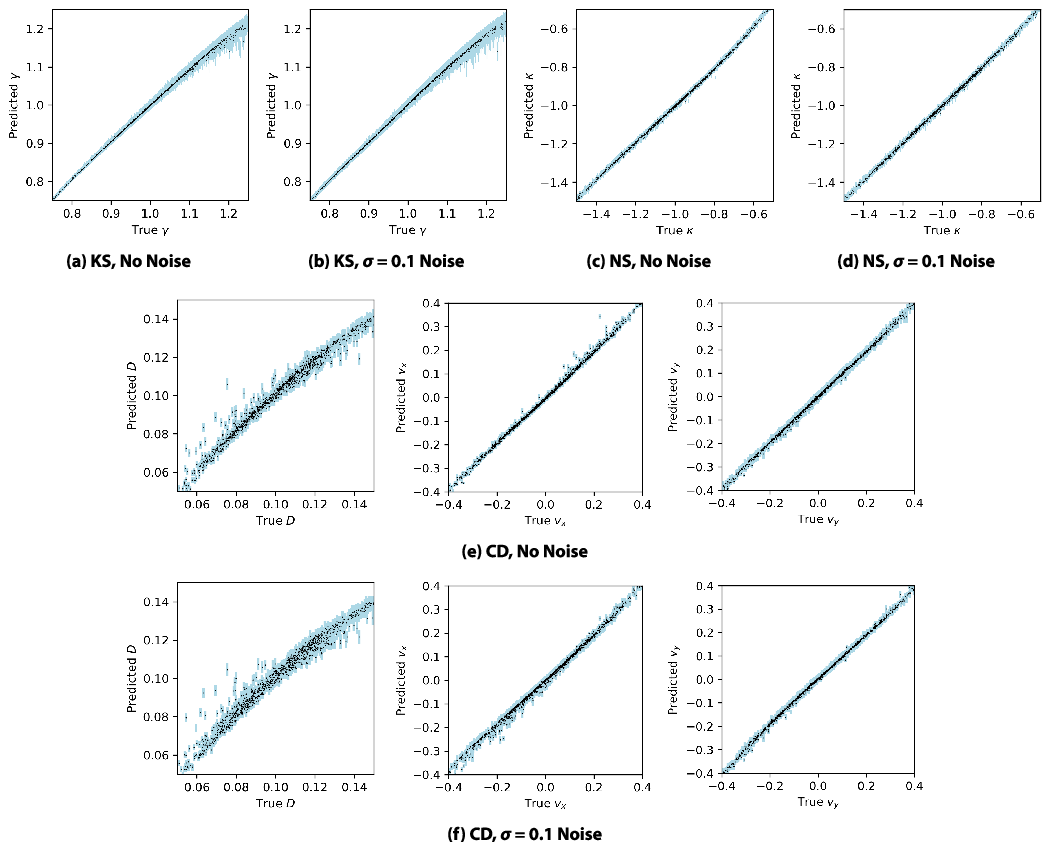}
    \caption{Predicted physical parameters from a linear fit with the relevant latent parameter (Fig.\ \ref{fig:relparams}) vs.\ the ground truth physical parameters from the (a), (b) 1D Kuramoto--Sivashinsky (KS), (c), (d) 1D nonlinear Schr\"odinger (NS), and (e), (f) 2D convection--diffusion (CD) datasets. Because the drift velocity $\mathbf{v}$ from the CD dataset has an inherent rotational symmetry, they are encoded in a two-dimensional latent subspace (Table \ref{tab:CD}), so we instead show the predicted drift velocity components $v_x$, $v_y$ from a multivariate linear regression in the subspace of two relevant latent parameters, which extracts the linear combination of latent parameters that correspond to $v_x$ and $v_y$. The light blue shaded bars are the 95\% confidence intervals produced by the models. Results are shown for models trained on (a), (c), (e) the original noiseless datasets as well as (b), (d), (f) the datasets with added $\sigma=0.1$ Gaussian noise.}
    \label{fig:paramfit}
\end{figure*}

During training, the model will only use a minimal set of latent parameters to encode the variation in the dynamics and will align each latent parameter in this subspace with an independent factor of variation due to the VAE regularization \cite{higgins2017beta,1804.03599}. Intuitively, the regularization encourages each latent parameter to independently collapse to a non-informative prior $\mathcal N(0,1)$, and so the model prefers to minimize its use of the latent parameters and maintain their independence (Appendix \ref{apx:vaereg}). Therefore, the number of latent parameters provided to the model is not critical as long as it is greater than the number of independent factors of variation. In our experiments, we allow the model to use five latent parameters. Because the 1D datasets have only one varying physical parameter and the 2D dataset has three varying physical parameters, the trained model will only make use of one or three latent parameters, respectively, and the rest will collapse to the prior.

We can determine the number of relevant latent parameters and empirically verify this claim by examining the statistics of the extracted distribution parameters $\mu_z$ and $\sigma_z^2$ from the dynamics encoder (DE) for each dataset. A latent parameter that is useful to the propagating decoder (PD) for predicting the target series will have high variance in $\mu_z$ and a low mean $\sigma_z^2$, implying that the extracted parameter is precise and informative. A parameter which has collapsed to the prior and is non-informative will have low variance in $\mu_z$ and high mean $\sigma_z^2$. These statistics indeed show that our model can correctly determine the number of relevant parameters for each dataset (Fig.\ \ref{fig:relparams}). In real applications, we will not have access to the ground truth physical parameters, so we must rely on these metrics to identify the relevant parameters extracted by the model.

To evaluate the performance of our parameter extraction method, we compare the extracted latent parameters from the model with the true physical parameters used to generate our simulated datasets: the viscosity damping parameter $\gamma$ for the 1D Kuramoto--Sivashinsky dataset, the nonlinearity parameter $\kappa$ for the 1D nonlinear Schr\"odinger dataset, and the diffusion constant $D$ and drift velocity components $v_x$, $v_y$ for the 2D convection--diffusion dataset. Because these simulated physical parameters are drawn from normal distributions (Appendix \ref{apx:datasets}), we expect the relevant latent parameters---which have prior distribution $\mathcal N(0,1)$---to be linearly related to the true parameters (Appendix \ref{apx:vaereg}). For real experimental systems, this is also a reasonable assumption for uncontrolled variables because natural parameters tend to be normally distributed due to the central limit theorem. We assess the quality of the extracted parameters by linearly fitting the relevant latent parameters with the ground truth physical parameters to obtain parameter predictions and $R^2$ correlation coefficients. Our numerical experiments show excellent parameter extraction on all three datasets (Fig.\ \ref{fig:paramfit}) with $R^2 > 0.95$ for all parameters (Table \ref{tab:corr}) and no degradation in performance with added Gaussian $\sigma=0.1$ noise. However, we do observe some nonlinear behavior at the edges of the parameter range likely due to data sparsity in those regions.

Looking more closely at the results for the three-parameter convection--diffusion dataset, we see that the trained model correctly extracts three relevant latent parameters: one latent parameter corresponds to the diffusion constant and the remaining two-dimensional latent subspace corresponds to the drift velocity vector (Table \ref{tab:CD}). In particular, the model learns a rotated representation of the drift velocity vector as a two-dimensional latent subspace due to the inherent rotational symmetry of the dynamics (Appendix \ref{apx:rawresults}), so we can recover the $v_x$, $v_y$ components of the drift velocity by performing a multivariate linear fit (Figs.\ \ref{fig:paramfit}(e), \ref{fig:paramfit}(f)). The successful separation of diffusion from drift velocity in the extracted parameters demonstrates our model's ability to distinguish distinct and interpretable factors of variation in the dynamics of the system.

\subsection{\label{sec:exp_pred}Prediction Performance}

\begin{figure*}
    \centering
    \includegraphics[width=0.95\linewidth]{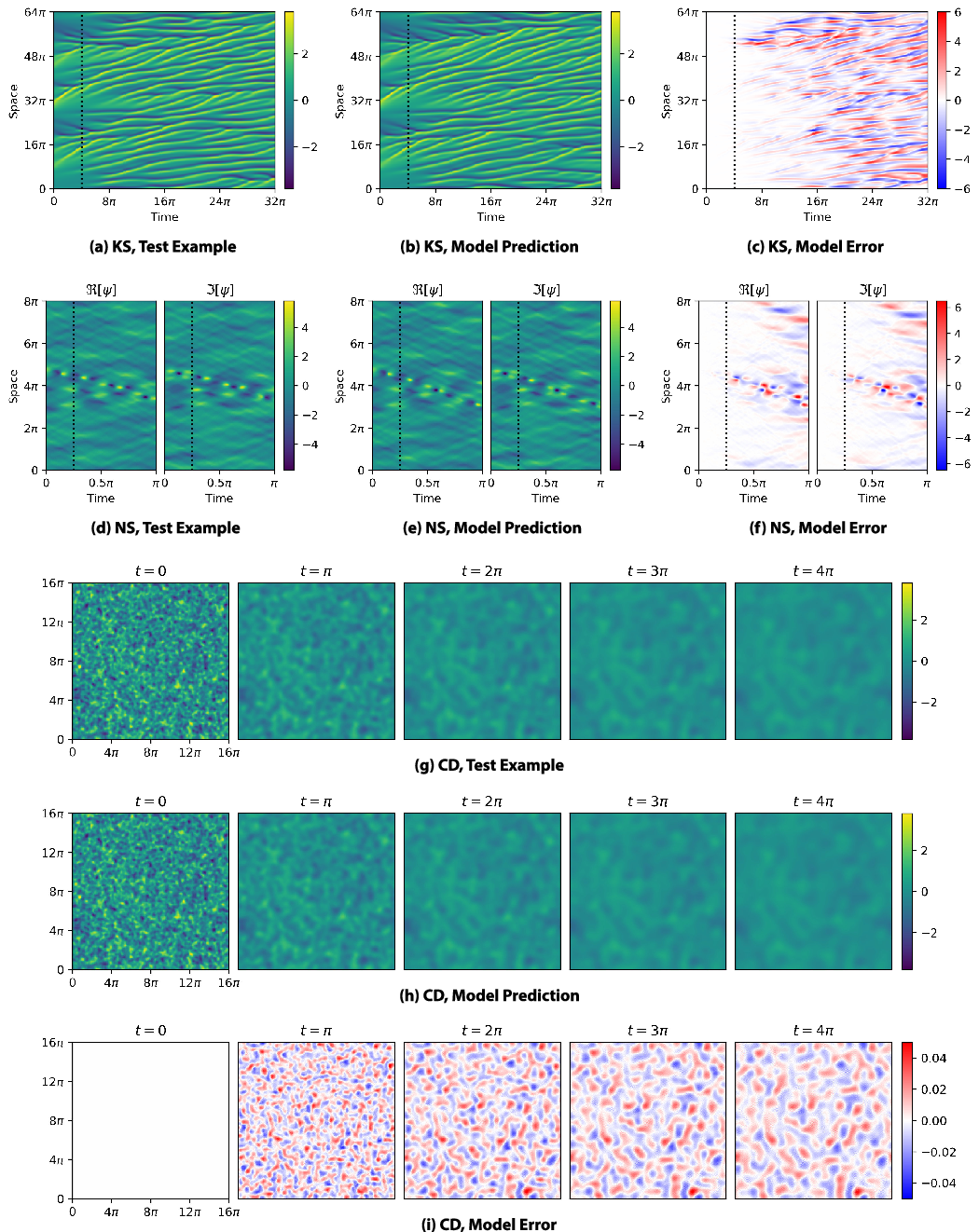}
    \caption{Time evolution of (a), (d), (g) examples from the 1D Kuramoto--Sivashinsky (KS), 1D nonlinear Schr\"odinger (NS), and 2D convection--diffusion (CD) test sets with periodic boundary conditions, (b), (e), (h) the predicted propagation of the refined model given the initial condition at time $t=0$, and (c), (f), (i) the model prediction error. The black vertical dotted line denotes the maximum amount of time propagated by the model during training, corresponding to the length of the target series. For the CD dataset, the maximum amount of time propagated by model during training is $t= \pi$.}
    \label{fig:examples}
\end{figure*}

\begin{figure*}
    \centering
    \includegraphics[width=\linewidth]{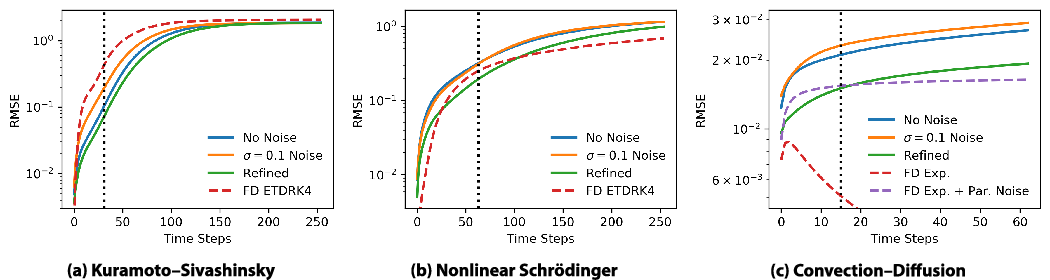}
    \caption{The root mean square prediction error (RMSE) during evaluation at each propagation time step, averaged over the corresponding test set, for models trained without noise (blue), trained with $\sigma=0.1$ Gaussian noise (orange), and refined by fixing the DE and training with 64 hidden channels in the PD network on the noiseless datasets (green). Shown for comparison is an evaluation by a second order finite difference discretization integrated with the ETDRK4 method \cite{kassam2005fourth} (dashed red), which reduces to an exact exponential integrator for the 2D convection--diffusion system. Also, since the prediction accuracy on 2D convection--diffusion is dominated by parameter extraction uncertainty, we include for comparison the solver with additional parameter noise: $\sigma=0.005$ for $D$, $\sigma = 0.01$ for $v_x$, $v_y$ (dashed purple). The black vertical dotted line denotes the length of the target series $T_y$ used for training each set of models, i.e.\ the maximum number of time steps propagated by the model during training.}
    \label{fig:pred}
\end{figure*}

In addition to testing parameter extraction, we evaluate the prediction performance of the trained models on their corresponding test sets. Due training speed and stability considerations, our models are initially trained with a PD architecture containing only 16 hidden channels (Appendix \ref{apx:model}). To show the potential for further model refinement, we fix the weights of the DEs trained on the original noiseless datasets and then train additional PDs, each with an expanded 64 hidden channels. These refined predictive models perform better than the original predictive models used during end-to-end training. For comparison, the datasets are also evaluated with a stiff exponential integrator for semilinear differential equations (ETDRK4 \cite{kassam2005fourth}) using a second order finite difference discretization on the same time and space meshes provided in the datasets. Although this integrator is the same one used during dataset generation, the time step and mesh size are set to match the available data to provide a reasonable baseline. During dataset generation, the solution is obtained starting with the exact form of the initial condition and converged using much finer mesh sizes (Appendix \ref{apx:datasets}). Also, note that using a non-stiff integrator fails on many of the examples in the datasets, so a stiff integrator is required.

The models trained on the 1D Kuramoto--Sivashinsky and 1D nonlinear Schr\"odinger datasets both perform reasonably when compared with the traditional finite difference method (Figs.\  \ref{fig:pred}(a), \ref{fig:pred}(b)), with the model trained on the Kuramoto--Sivashinsky dataset maintaining a higher accuracy than its traditional counterpart. The prediction error of the 2D convection--diffusion model is dominated by the uncertainty in the parameter extraction, so the prediction performance is comparable to a finite difference exponential integrator with similar noise in the PDE parameters (Fig.\ \ref{fig:pred}(c)). For the models trained on datasets with added $\sigma=0.1$ noise, we see some negative impact on prediction performance (Fig.\ \ref{fig:pred}) but no effect on the parameter extraction quality (Tables \ref{tab:corr}, \ref{tab:CD}).

The refined models, trained on the noiseless datasets, demonstrate that the PD---the predictive network---can be improved independently of the DE---the parameter extraction network. Moreover, the solutions generated by these models remain stable and physically reasonable well beyond the number of time steps propagated during training, suggesting that the models have indeed learned physically meaningful propagators of the PDE-governed systems (Fig.\ \ref{fig:examples}).

\section{\label{sec:fiber}Application to Nonlinear Fiber Propagation}

\begin{figure*}
    \centering
    \includegraphics[width=\linewidth]{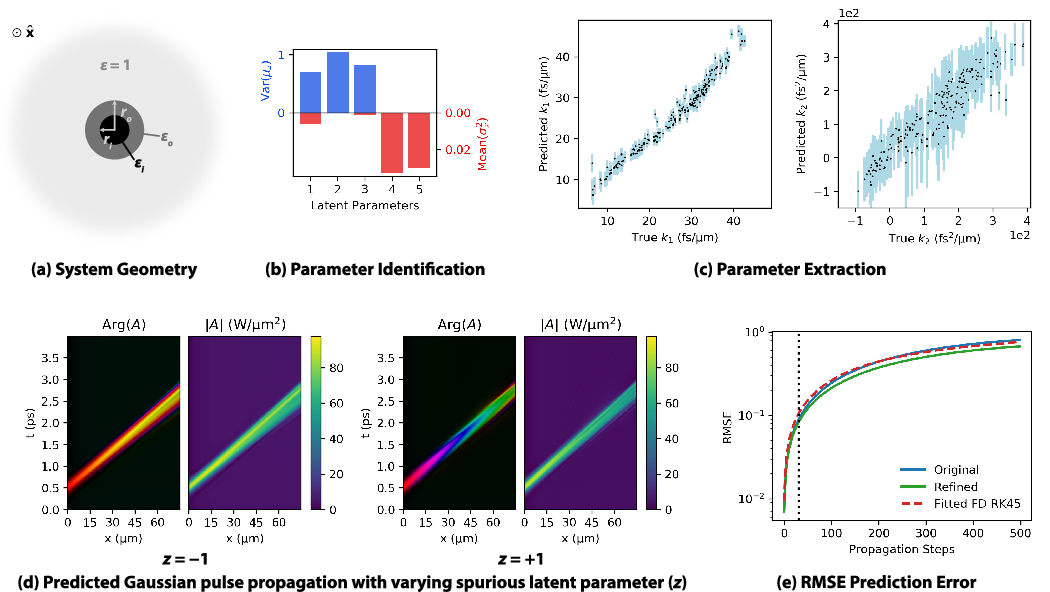}
    \caption{Analysis of the model trained on the nonlinear fiber propagation dataset. (a) The cross section of the fiber shows the cladding with (relative) permittivity $\varepsilon = 1$, the inner core with radius $r_i$ and permittivity $\varepsilon_i$, and the outer core with radius $r_o$ and permittivity $\varepsilon_o$. (b) Three relevant latent parameters are identified by the trained model. (c) A linear fit of the latent parameters predicts the true $k_1$ with $R^2 = 0.966$ and $k_2$ with $R^2 = 0.863$, corresponding to group velocity and second-order dispersion. (d) Varying the remaining spurious latent parameter $z$ while propagating a Gaussian pulse using the trained propagating decoder (PD), we observe that the spurious parameter represents a phase velocity. The plot of $\mathrm{Arg}(A)$ shows the phase of the pulse mapped to the color hue while the amplitude is mapped to lightness. (e) The root mean squared prediction error (RMSE) at each propagation step of the originally trained PD (blue) and a further refined PD (green) are comparable to an explicit solution (second order finite difference, adaptive fourth order Runge--Kutta) of the effective equation (\ref{eqn:fiber}) with fitted parameters (up to fourth-order dispersion) for each individual test set example (dashed red).}
    \label{fig:fiber}
\end{figure*}

To demonstrate our method applied on a more complex and realistic dataset, we use Meep \cite{OSKOOI2010687}, a finite difference time-domain electromagnetic simulator, to model pulse propagation through an optical fiber with a Kerr nonlinearity. These simulations model Maxwell's equations exactly, with no approximation apart from the discretization, and often reproduce real experiments point-by-point \cite{taflove2005computational}. The simulated fiber consists of a two-layer core and the surrounding cladding with (relative) permittivity $\varepsilon = 1$ (cross section shown in Fig.\ \ref{fig:fiber}a). We generate a dataset with 200 different geometries by varying the size of the two-layer core through $r_i$ and $r_o$ as well as the corresponding permittivities $\varepsilon_i$ and $\varepsilon_o$, representing uncontrolled experimental variables related to fabrication (see Appendix \ref{apx:datasets} for details). Then, we excite a randomly generated pulse in each fiber and train our model on the flux-normalized amplitude $A(x,t)$ of the resulting pulse propagation.

There is no exact first-order PDE describing the evolution of this amplitude. However, in the slowly varying envelope approximation, $A(x,t)$ is governed by an effective nonlinear Schr\"odinger equation \footnote{This effective equation is usually transformed to a co-moving time coordinate $\tau = t- x/v_g$, where $v_g$ is the group velocity \cite{BOYD2008329}. However, since the group velocity varies among of the dataset examples and because we want to use minimal preprocessing, we keep the original time coordinate $t$. We also include higher order dispersion terms which are relevant for our dataset.}
\begin{equation}
    \frac{\partial A}{\partial x} =  i\sum_{n=1}^\infty\frac{i^nk_n}{n!}\partial_t^nA + i\gamma \left|A\right|^2A,
    \label{eqn:fiber}
\end{equation}
where the dispersion coefficients $k_n = \partial^n_\omega k(\omega)|_{\omega =\omega_0}$ can be computed from the dispersion relation $\omega(k)$ at a carrier frequency $2\pi\omega_0$, and the nonlinearity parameter $\gamma$ is related to the Kerr nonlinearity (Appendix \ref{apx:datasets}) and the shape of the propagating mode in the fiber \cite{BOYD2008329}. Due to the form of this effective equation, we choose our propagation variable to be the distance $x$ rather than the time $t$, i.e.\ our model will predict $A(x,t)$ given an initial pulse $A_0(x=0,t)$ by propagating forward in the $x$-direction. We obtain the ground truth dispersion coefficients using MPB \cite{Johnson:01}, a frequency domain electromagnetic eigenmode solver, for comparison with the extracted parameters from the model.

The trained model identified and extracted three relevant latent parameters (Fig.\ \ref{fig:fiber}b). Two independent directions in the latent parameter space (primarily, latent parameters 2 and 3) correspond to the group velocity $k_1 = 1/v_g$ with $R^2 = 0.966$ and second-order dispersion $k_2$ with $R^2 = 0.863$ (Fig.\ \ref{fig:fiber}c). Note that the extracted parameters are not the geometry parameters used to generate the dataset, but rather parameters relevant to the effective propagation of the pulse. These two latent parameters also capture variations in higher order dispersion terms, which are correlated with the group velocity and second-order dispersion in the dataset. As a result, although important and present in the effective equation (\ref{eqn:fiber}), the higher order terms are already well correlated with the existing latent parameters, so no additional parameters are required to characterize the dynamics. See Appendix \ref{apx:fiberanalysis} for more parameter analysis details.

In addition to the parameters corresponding to $k_1$ and $k_2$, the model extracted another relevant latent direction (primarily, latent parameter 1), which is independent and orthogonal to the previous two. This seemingly spurious latent parameter does not correspond to a term in the effective equation (\ref{eqn:fiber}); instead, it represents a spurious phase velocity, which is the result of imperfect preprocessing (Appendix \ref{apx:datasets}). We discovered this correspondence by varying the spurious parameter, while leaving all other latent parameters fixed, and observing the effect on the model predictions (Fig.\ \ref{fig:fiber}d). This uncontrolled variable was successfully extracted by the model and subsequently identified, demonstrating the process by which unknown extracted parameters can be understood and interpreted.

We evaluate the prediction performance of the trained propagating decoder (PD) as well as a refined version of the PD (with an expanded 64 hidden channels) by comparing the prediction error of our models with an explicit solution to the effective equation (\ref{eqn:fiber}) as a baseline. The effective equation parameters are determined by a linear fit of the finite difference derivatives computed from the dataset. For each example in the test set, we fit the dispersion parameters $k_1$, $k_2$, $k_3$, $k_4$ (i.e.\ up to fourth order dispersion) and the nonlinearity parameter $\gamma$. The effective equation is then integrated using a fourth order Runge--Kutta method with adaptive step size and second order finite difference discretization. This approach is a simplified version (with the terms pre-identified) of explicit PDE identification methods, such as SINDy \cite{Rudy:2017cs}. Our predictive models achieve similar prediction performance when compared with this explicit effective equation baseline (Fig.\ \ref{fig:fiber}e).

\section{\label{sec:dis}Discussion}

We have developed a general unsupervised learning method for extracting unknown dynamical parameters from noisy spatiotemporal data {\em without} detailed knowledge of the underlying system or the exact form of the governing PDE. While we do not explicitly extract the governing PDE, our method provides a set of tunable relevant parameters, which characterize the system dynamics in independent and physically interpretable ways, coupled with a highly transferable predictive model. This is often enough to provide significant insight into the physics of the system: for example, by examining the effect of varying an unidentified relevant parameter on the predictions of the model, we can disentangle the parameter from other effects (parameterized by the other relevant parameters) and interpret it independently. This is precisely how we identified the spurious phase velocity parameter extracted by the model trained on nonlinear fiber propagation (Fig.\ \ref{fig:fiber}d). One potential complication of interpreting the parameters extracted using our method is that each parameter must represents an independent factor of variation in the dynamics of the dataset. This means that if features of the dynamics are highly correlated in the underlying dataset, they will be parameterized by the same parameter, e.g.\ the parameters extracted from the nonlinear fiber propagation dataset corresponding to group velocity and second order dispersion also capture higher order dispersion terms (Appendix \ref{apx:fiberanalysis}).

The flexibility and robustness of our model comes from using a generic physics-informed neural network model for nonlinear PDEs. The interpretability of the resulting extracted parameters is a result of the variational autoencoder (VAE) training and regularization (Appendix \ref{apx:vaereg}) as well as the inductive biases imposed by our physics-informed network design. By using appropriate spatial averaging in the dynamics encoder (DE) and dynamic convolutions in the propagating decoder (PD), we ensure that both the parameter extraction and the propagation prediction from our model are physically motivated and generalizable to arbitrary system sizes and geometries. The dynamic convolutions, in particular, are an important physical inductive bias for encouraging the model to learn latent parameters which govern the propagation dynamics. As a result, the learned parameter-to-kernel mappings in the trained predictive model are fully transferable, which we can demonstrate by evaluating the predictive model {\em without retraining} on a different set boundary conditions (see Appendix \ref{apx:altboundary}).

Our strategy for modeling spatiotemporal systems is to retain the expressiveness of a neural network model while imposing general constraints---such as locality through using convolutional layers---to help the network learn more efficiently. For particular applications, this could also include spatial symmetries \cite{pmlr-v48-cohenc16,cohen2017steerable,NIPS2018_8239}, e.g.\ properly transforming fluid flow vectors, as well as additional symmetries of the internal degrees of freedom, e.g.\ the global phase of the nonlinear Schr\"odinger equation. These architecture-based constraints encourage the model to learn physically relevant representations and can be tailored to individual applications, allowing us to incorporate domain knowledge into the model. This also lets us use datasets that are much smaller than is traditionally required by deep learning methods. The model trained on nonlinear fiber propagation used only 200 examples (Sec.\ \ref{sec:fiber}), and we have been able to successfully train models on the Kuramoto--Sivashinsky dataset with as few as 10 examples (Appendix \ref{apx:datasize}). Additionally, the model's robustness to noise is a powerful feature of deep learning methods and provides a promising avenue for studying dynamical systems in a data-driven fashion \cite{RUDY2019}.

The primary challenges associated with applying our current implementation involve setting hyperparameters to improve training stability (see Appendix \ref{apx:stability} for more details) and choosing the $\beta$ regularization hyperparameter, which controls parameter extraction (Appendix \ref{apx:vaereg}). The choice of $\beta$ can be somewhat ambiguous, with very high values resulting in no relevant parameters and very low values failing to enforce independence (for our choices, see Appendix \ref{apx:train}). This trade-off and other related issues are a very active area of research, and parameter extraction methods will continue to improve following the rapid advances in unsupervised learning and disentangling representations, e.g.\ through a deeper theoretical understanding of the $\beta$-VAE \cite{alemi18a,1804.03599,1702.08658} and alternative formulations \cite{NIPS2018_7527,infoVAE}. Physics-informed inductive biases, however, will remain the key ingredient for ensuring the representations are interpretable \cite{pmlr-v97-locatello19a}.

The predictive model will also likely achieve better accuracies by using more sophisticated architectures, such as echo state networks which have been shown to perform extremely well on even chaotic PDEs \cite{PhysRevLett.120.024102}, or by explicitly incorporating differentiable PDE solvers with gradients computed by the adjoint method \cite{NIPS2018_7892}. An echo state network or other alternative decoder architecture would have to be adapted to retain the transferability of our current PD. Using a differentiable PDE solver would allow the PD network to focus on encoding the PDE rather than also learning a stable integration method, and thus may improve the interpretability of the model.

Our unsupervised learning method is also highly complementary with the significant body of work applying machine learning methods for more accurate predictions of specific physical systems, such as multi-scale hydrodynamic systems \cite{Han21983} and turbulence modeling \cite{CiCP-25-947,ling_kurzawski_templeton_2016,PhysRevFluids.3.074602,PhysRevFluids.4.034602}. These methods often combine a known physics model with a machine learned correction or parameterize an unknown part of the physics model using neural networks. Such machine learning--physics hybrid models can be adapted into decoders for our unsupervised learning method, with the extracted relevant parameters representing independent variations of and providing insight into the machine learned portions of the model. Our current method can also be adapted in the future for a more general class of spatiotemporal systems by incorporating spatially inhomogeneous latent parameters and will also be able to use data with incomplete physical state observations by inferring missing information.

The ultimate goal of this work is to provide additional insight into complex spatiotemporal dynamics using a data-driven approach. Our method is an example of a new machine learning tool for studying the physics of spatiotemporal systems with an emphasis on interpretability.

\begin{acknowledgments}
We would like to acknowledge useful discussions with Jason Fleischer, Rumen Dangovski, and Li Jing. This research is sponsored in part by the Army Research Office and was accomplished under Cooperative Agreement Number W911NF-18-2-0048. This work is further supported in part by the U.S. Department of Defense through the National Defense Science \& Engineering Graduate Fellowship (NDSEG) Program and by the MIT--SenseTime Alliance on Artificial Intelligence. This material is also based in part upon work supported by the Defense Advanced Research Projects Agency (DARPA) under Agreement No.\ HR00111890042.
\end{acknowledgments}

\appendix

\section{Model Implementation Details\label{apx:model}}

For the 1D datasets, the dynamics encoder (DE) uses 2D convolutions with output channel sizes $(4, 16, 64, 64, 5)$, linear kernel sizes $(3, 3, 3, 3, 1)$, and dilation factors $(1, 2, 4, 8, 1)$. For the 2D dataset, the DE uses 3D convolutions with output channel sizes $(8, 64, 64, 64, 5)$ with the same kernel sizes and dilation factors. The $f_{\mathrm{DE},\mu}$ and $f_{\mathrm{DE},\sigma^2}$ networks share the same convolution weights for the first four layers and have distinct final layers to produce $\mu$ and $\log\sigma^2$. The final output channel size is determined by the number of latent parameters; in our tests, we use five latent parameters.

For the 1D datasets, the propagating decoder (PD) architecture uses three 1D dynamic convolutional layers with output channel sizes $(16, 16, \mathrm{data\ channel\ size})$, linear kernel size $5$, and periodic padding. For the 2D datasets, the PD uses three 2D dynamic convolutional layers with the same output channel sizes, kernels, and padding. The refined models increase the number of hidden channels in the PD from 16 to 64, resulting in output channel sizes $(64, 64, \mathrm{data\ channel\ size})$.

The latent-to-kernel network, which maps the latent parameters to each kernel or bias in the PD, consists of two fully-connected layers, i.e.\ one hidden layer. For the 1D datasets, the hidden layers have size $(4\times\mathrm{input\ channel\ size}\times \mathrm{output\ channel\ size})$ for kernels and $(4\times \mathrm{output\ channel\ size})$ for biases, where the input and output channel sizes refer to the corresponding dynamic convolution in the PD. For the 2D dataset, the hidden layers have size $(16\times\mathrm{input\ channel\ size}\times \mathrm{output\ channel\ size})$ for kernels and $(4\times \mathrm{output\ channel\ size})$ for biases.

Our architecture uses ReLU activations throughout except for the unactivated output layers of the DE, PD, and latent-to-kernel networks. The output of the PD convolutional network $f_\mathrm{PD}$ uses a tanh activation with a learnable scaling parameter ($x \mapsto \lambda \tanh(x/\lambda)$ with learnable parameter $\lambda$ initialized to 1) to stabilize the recurrent architecture. The network $f_\mathrm{PD}$ also has a fixed multiplicative pre-factor set to $10^{-6}$ to improve the initial training stability. For the nonlinear fiber propagation dataset, we add Gaussian noise with $\sigma = 10^{-2}$ between propagation steps in the PD network for improved prediction stability.

Our model is implemented using PyTorch v1.1 \cite{NIPS2019_9015}, and the source code is available at \url{https://github.com/peterparity/PDE-VAE-pytorch}.

\section{Training Details\label{apx:train}}

All models are trained using batch size 50 and the Adam optimizer \cite{kingma2015adam} with learning rate $10^{-3}$. Models for the 1D datasets and the noisy 2D convection--diffusion dataset were trained for 2,000 epochs; the model for the noiseless 2D convection--diffusion dataset was trained for 4,000 epochs; and the corresponding refined models were trained for 2,000 epochs. The VAE regularization hyperparameter is set to $\beta=0.02$ for the 1D datasets and $\beta=10^{-4}$ for the 2D convection--diffusion dataset. The model for the nonlinear fiber propagation application was trained for 40,000 epochs due to small size of the dataset and significant data augmentation; the corresponding refined model was trained for 20,000 epochs; and the VAE regularization hyperparameter was set to $\beta = 7\times 10^{-4}$. During validation, we choose $\beta$ such that we obtain the maximum number of relevant latent parameters while still maintaining statistical independence among the parameters as well as a clean separation between the relevant and irrelevant (i.e.\ collapsed to the prior) parameters. All hyperparameter tuning is done using the training set for validation.

For the 1D Kuramoto--Sivashinsky dataset, we train the model using a random $64\times94$ crop---in the time and space dimensions, respectively---for the input series and another random $64\times 76$ crop for the target series. For the 1D nonlinear Schr\"odinger dataset, we train using a $64\times94$ crop for the input series and a $32\times 76$ crop for the target series. For the 2D convection--diffusion dataset, we train using a $45\times62\times 62$ crop---in the one time and two space dimensions, respectively---for the input series and a $16\times 44 \times 44$ crop for the target series. For the nonlinear fiber propagation dataset, we train using a $128\times158$ crop for the input series and a $32\times 76$ crop for the target series. During evaluation, we always use the full size time-series from the test set for both the input and target series.

\section{Notes on Training Stability\label{apx:stability}}

We find our model to be particularly sensitive to the architecture of the propagating decoder (PD): with larger, more complex networks and more propagation steps during training resulting in increasing instability. The dynamics encoder (DE) does influence stability, but the effect is more indirect through its interaction with the PD and is not very architecture sensitive. This instability is likely related to the problem of vanishing and exploding gradients seen often in recurrent architectures, which is mitigated using gating mechanisms like in LSTM networks \cite{hochreiter1997long,doi:10.1162/089976600300015015} or by explicitly using unitary norm-preserving matrices \cite{jing17a}. Importantly, this is only a significant problem when training the model end-to-end using both the DE and PD; when we fix the DE weights, we are able to further refine a more complex PD model without instability (Sec.\ \ref{sec:exp_pred}). This also does {\em not} affect the prediction performance and stability of the model during evaluation, which generalizes well past the number of time steps propagated during training (Fig.\ \ref{fig:pred}). We currently implement a learnable gating mechanism (Appendix \ref{apx:model}) that significantly stabilizes the network but further work is required to fully address this issue.

\section{Dataset Generation Details\label{apx:datasets}}

For each time-series example in the 1D Kuramoto--Sivashinsky dataset, we sample the viscosity damping parameter $\gamma$ from a truncated normal distribution ($\mu = 1$, $\sigma = 0.125$, and truncation interval $[0.5, 1.5]$). We then use the ETDRK4 integrator \cite{kassam2005fourth} to generate each time-series to within a local relative error of $10^{-3}$. Each time-series consists of a uniform time mesh with $256$ points for a total time $T = 32\pi$ and a space mesh with $M = 256$ points for an $L = 64\pi$ unit cell. These are produced by solving on a finer time and space mesh to ensure convergence and then resampling to the dataset mesh sizes. Each initial state is generated from independently sampled, normally distributed Fourier components with a Gaussian band-limiting envelope of varying widths (uniformly sampled in the interval $[8\pi/L, M\pi/4L]$) and then normalized to unit variance.

For each time-series example in the 1D nonlinear Schr\"odinger, we sample the nonlinearity coefficient $\kappa$ from a truncated normal distribution ($\mu = -1$, $\sigma = 0.25$, and truncation interval $[-2, 0]$). We then use the ETDRK4 integrator \cite{kassam2005fourth} to generate each time-series to within a local relative error of $10^{-3}$. Each time-series consists of a uniform time mesh with $256$ points for a total time $T = \pi$ and a space mesh with $M = 256$ points for an $L = 8\pi$ unit cell. These are produced by solving on a finer time and space mesh to ensure convergence and then resampling to the dataset mesh sizes. The initial states are generated in an analogous manner to the Kuramoto--Sivashinsky dataset.

For each time-series example in the 2D convection--diffusion dataset, we vary the parameters by sampling the diffusion constant $D$ from a truncated normal distribution ($\mu = 0.1$, $\sigma = 0.025$, and truncation interval $[0, 0.2]$), and each velocity component $v_x$, $v_y$ from a normal distribution ($\mu = 0$, $\sigma = 0.2$). Because the convection--diffusion equation is linear, we use the exact solution to generate the dataset. Each time-series consists of a uniform time mesh with $64$ points for a total time $T = 4\pi$ and an $M \times M = 256 \times 256$ space mesh for an $L \times L = 16\pi \times 16\pi$ unit cell. Each initial state is generated from independently sampled, normally distributed Fourier components with a Gaussian band-limiting envelope of varying widths (uniformly sampled in the interval $[16\pi/L, M\pi/2L]$) and then normalized to unit variance.

The nonlinear fiber propagation dataset aims to roughly model a set of highly dispersive nonlinear optical fibers with variations in geometry due to fabrication. Each of the 200 examples in the nonlinear fiber propagation dataset corresponds to a randomly generated fiber geometry excited with a randomly generated pulse. Specifically, each fiber geometry (Fig.\ \ref{fig:fiber}a) corresponds to a set of geometry parameters sampled from normal distributions: inner core radius $r_i$ ($\mu = \SI{75}{\nano\meter}$, $\sigma = \SI{3}{\nano\meter}$), outer core radius $r_o$ ($\mu = \SI{150}{\nano\meter}$, $\sigma = \SI{7.5}{\nano\meter}$), inner core (relative) permittivity $\varepsilon_i$ ($\mu = 30$, $\sigma = 2$), and outer core permittivity $\varepsilon_o$ ($\mu = 8$, $\sigma = 1$). There is an overall fixed Kerr nonlinearity that corresponds to a nonlinear refractive index $n_2 = (\SI{3.375e-4}{\micro\meter\squared\per\watt})/\varepsilon$, where $\varepsilon$ is the material permittivity. The excited pulse is generated from independently sampled, normally distributed frequency components with a Gaussian band-limiting envelope centered on a carrier frequency $f = \SI{200}{\tera\hertz}$ ($\lambda = \SI{1.5}{\micro\meter}$) and with a width of $f/20$. This pulse is slowly turned on with a sigmoid of width $20/f$, and, for the test set, the pulse is also turned off at half the total simulation time, allowing the pulse to fully propagate through the fiber. During the simulation, the amplitude $A(x,t)$ of the electric field at the center of fiber is recorded and later normalized to the total flux passing through the fiber. The space and time Fourier components of each resulting dataset example $A(x,t)$ are shifted to remove the carrier frequency and the peak wavenumber, resulting in a slowly varying envelope. Then, the final dataset is normalized again so that the amplitude at the initial point $x=0$ has, on average, unit variance over the whole dataset. Each example consists of an $x$-direction mesh with $500$ points for a propagation length of $\SI{75}{\micro\meter}$ and a uniform $t$-direction mesh with $800$ points for a total time $\SI{4.00}{\pico\second}$ (training set) or $1000$ points for a total time $\SI{5.00}{\pico\second}$ (test set).

The dataset generation scripts are available at \url{https://github.com/peterparity/PDE-VAE-pytorch}.

\section{Understanding the Effects of VAE Regularization\label{apx:vaereg}}

Using VAE \cite{1312.6114} or $\beta$-VAE \cite{higgins2017beta} regularization in our model provides three main benefits for learning physically interpretable representations: the regularization encourages the model to minimize use of the latent parameters, enforces independence among the learned latent parameters, and matches the marginal latent distribution to a standard normal prior. We can explicitly see these effects by decomposing the data-averaged VAE regularization term in the following way \cite{NIPS2018_7527,hoffman2016elbo}:
\begin{align}
    \mathbb E_{p_D(\mathbf{x})}[D_\mathrm{KL}&(q(\mathbf{z}|\mathbf{x})\;\|\;p(\mathbf{z}))] = \nonumber\\
    & D_\mathrm{KL}(q(\mathbf{z}, \mathbf{x})\;\|\;q(\mathbf{z})\,p_D(\mathbf{x})) \label{eqn:vaereg1}\\
    & + D_\mathrm{KL}(q(\mathbf{z})\;\|\;\prod_iq(z_i)) \label{eqn:vaereg2}\\
    & + \sum_i D_\mathrm{KL}(q(z_i)\;\|\;p(z_i)), \label{eqn:vaereg3}
\end{align}
where $p_D(\mathbf{x})$ is the data distribution, $p(\mathbf{z}) = \prod_i p(z_i)$ are the standard normal priors for the latent parameters $\mathbf{z} = (z_1, z_2, \ldots, z_i, \ldots)$, $q(\mathbf{z}|\mathbf{x})$ is the output distribution of the dynamics encoder, $q(\mathbf{z},\mathbf{x}) = q(\mathbf{z}|\mathbf{x})\,p_D(\mathbf{x})$ is the joint distribution of the encoded latent parameters and the data, and $q(\mathbf{z}) = \int d\mathbf{x}\,q(\mathbf{z},\mathbf{x})$ and $q(z_i) = \int d\mathbf{x}\,\prod_{j\ne i}dz_j\,q(\mathbf{z},\mathbf{x})$ are the marginal distributions of the latent parameters $\mathbf{z}$ or a single latent parameter $z_i$, respectively. The three terms in this decomposition correspond directly to the three effects: the first term (\ref{eqn:vaereg1}) represents the mutual information between the latent parameters and the data; the second term (\ref{eqn:vaereg2}) represents the total correlation between the latent parameters; and the third term (\ref{eqn:vaereg3}) consists of KL divergences between the marginal distribution for individual latent parameters and the standard normal prior.

By minimizing the mutual information between the latent space and the data (\ref{eqn:vaereg1}) as well as correlations among the latent parameters (\ref{eqn:vaereg2}), the model is compelled to learn a latent space with minimal information and independent parameters, i.e.\ the model will use a minimal set of independent relevant latent parameters to capture only the necessary information for better prediction performance. The rest of the unused latent parameters will collapse to the prior. Furthermore, by matching the {\em marginal} latent parameter distributions $q(z_i)$ to the standard normal priors $p(z_i)$ (\ref{eqn:vaereg3}), the VAE regularizer encourages a linear relationship between the relevant learned latent parameters and the true physical parameters {\em if} the physical parameters are normally distributed in the data. Even if a physical parameter $z_\mathrm{phys}$ is non-normally distributed, the VAE regularization will still compel the model to learn a monotonic relationship between $z_\mathrm{phys}$ and a corresponding latent parameter $z$ given by
\begin{equation}
    z_\mathrm{phys} = \pm\,\mathrm{CDF}_{p(z_\mathrm{phys})}\circ \mathrm{CDF}_{p(z)}^{-1}(z),
\end{equation}
where $\mathrm{CDF}_{p(\cdot)}$ is the cumulative distribution function for the probability distribution $p(\cdot)$. One caveat---in addition to ambiguities introduced by symmetries of the physical parameters---is that the relationship may not be monotonic for physical parameter distributions which have support on a topologically distinct space from the real line, e.g.\  a uniformly distributed periodic angle parameter. However, the result may still be interpretable, e.g.\ an angle parameter may be encoded as a ring in a two-dimensional latent subspace.

Although this decomposition is suggestive of the effects of VAE regularization, the study of the performance of VAE-based models and the relative importance and model dependence of each of these effects is still very much ongoing \cite{alemi18a,1804.03599,NIPS2018_7527,infoVAE,1702.08658,1802.06847}. While training our model, we empirically observe that the latent parameters retain their independence and that their marginal distributions match the standard normal priors, so only an increase in information stored in the latent space is traded for better prediction performance. We believe we can attribute this to the physics-informed inductive biases present in our architecture, which allows our model to achieve its best performance using a minimal set of independent and normally distributed latent parameters.

\begin{figure*}
    \centering
    \includegraphics[width=\linewidth]{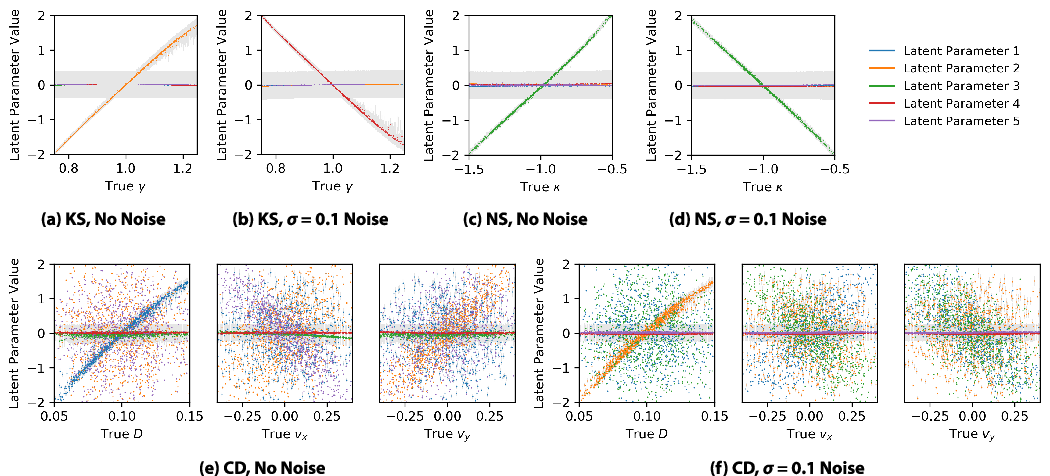}
    \caption{The five latent parameters in the models trained on the (a), (b) 1D Kuramoto--Sivashinsky (KS), (c), (d) 1D nonlinear Schr\"odinger (NS), and (e), (f) 2D convection--diffusion (CD) datasets both with and without added noise vs.\ the ground truth physical parameters used to generate the datasets. In these plots of the raw extracted latent parameters, we see the direct correspondence between the identified relevant latent parameters in Fig.\ \ref{fig:relparams} and the true physical parameters as well as the collapse of the unused latent parameters. Note that the relevant latent parameters corresponding the drift velocity $\mathbf{v}$ in the convection--diffusion model are not precisely aligned with the two components $v_x$, $v_y$ due to the inherent rotational symmetry. The gray shaded bars are the 95\% confidence intervals ($\pm 1.96\sigma_z$) produced by the model.}
    \label{fig:rawplots}
\end{figure*}

\begin{figure*}
    \centering
    \includegraphics[width=\linewidth]{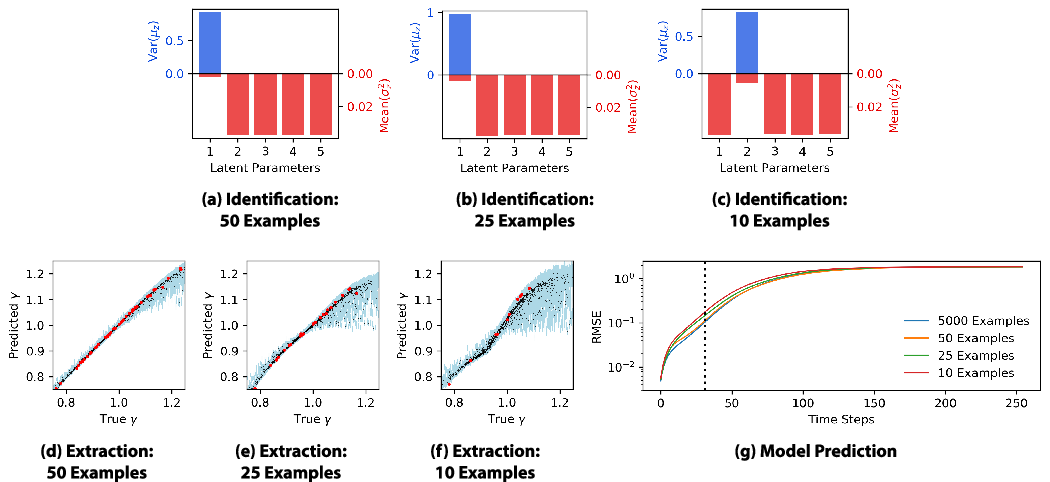}
    \caption{Parameter extraction and prediction performance for models trained on noiseless 1D Kuramoto--Sivashinsky datasets with 50, 25, and 10 examples to show the effect of dataset size. The (a), (b), (c) parameter identification and (d), (e), (f) extraction plots demonstrate the ability of the model to identify and extract a relevant latent parameter using very few examples. In the extraction plots, the small black points show the evaluation on the test set, the large red points show the training examples used for each model, and the light blue shaded bars are the 95\% confidence intervals produced by the model. There is also a subtle decrease in prediction performance seen in (g) the root mean square prediction errors (RMSE) averaged over the 10,000 example test set. The parameter identification and extraction plots for the 5,000 example dataset are shown in Figs.\  \ref{fig:relparams}(a) and \ref{fig:paramfit}(a), respectively.}
    \label{fig:datasizetest}
\end{figure*}

\begin{figure*}
    \centering
    \includegraphics[width=\linewidth]{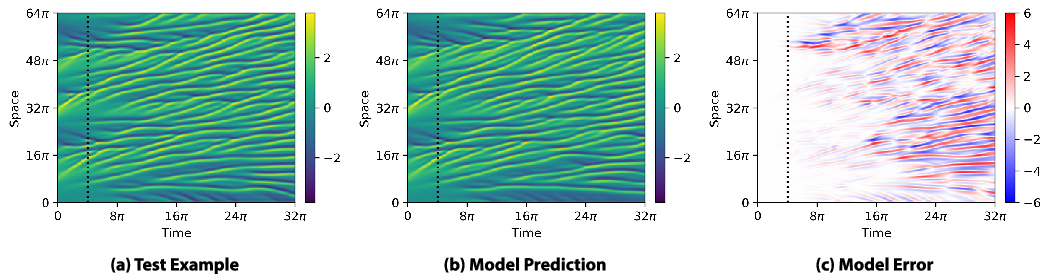}
    \caption{Time evolution of (a) a 1D Kuramoto--Sivashinsky example with Dirichlet boundary conditions, (b) the predicted propagation using transferred kernels from the refined predictive model given the initial condition at time $t=0$, and (c) the prediction error of the model. The refined predictive model for the 1D Kuramoto--Sivashinsky was originally trained on data with periodic boundaries and is adapted---without retraining---to use Dirichlet hard wall boundaries instead by adjusting the padding at each propagation time step. The black vertical dotted line denotes the maximum amount of time propagated by the model during the original training, corresponding to the length of the target series.}
    \label{fig:KS7dirichlet}
\end{figure*}

\section{Raw Parameter Extraction Results\label{apx:rawresults}}

We can explicitly see the relevant and collapsed latent parameters in the raw data by plotting the latent parameters versus the true physical parameters (Fig.\ \ref{fig:rawplots}). The latent parameters that show a correlation with the true physical parameters also have small variances $\sigma_z^2$ and correspond to the identified relevant latent parameters (Fig.\ \ref{fig:relparams}), while the remaining latent parameters have collapsed to the prior. Note that for the collapsed parameters, we see variances $\sigma_z^2$ that are less than one---the value expected for parameters which have collapsed to the prior distribution $\mathcal N(0,1)$. This is because, to average over systems of different sizes, the model makes the assumption that patches separated far in space or time provide independent estimates of the extracted parameters and computes the total variance accordingly. This assumption is reasonable for relevant parameters but will artificially lower the extracted variances for collapsed parameters. During testing, we choose to evaluate on the full system size resulting in this artifact. If we were to evaluate on smaller patches that match the size of the crops used during training, we would indeed see that the collapsed parameters have $\sigma_z^2=1$.

We also note that, for the model trained on the 2D convection--diffusion dataset, the latent parameters associated with the drift velocity $\mathbf{v}$ are not aligned with the $v_x$, $v_y$ velocity components. This is an expected result due to the inherent ambiguity of choosing a coordinate basis---introduced by the rotational symmetry of the velocity vector---and makes judging the extraction performance more difficult. Instead of examining one latent parameter at a time, we must consider the two-dimensional latent subspace associated with the velocity vector. Taking the two relevant latent parameters that are correlated with the drift velocity (Table \ref{tab:CD}), we can perform a multivariate linear regression of the velocity components $v_x$,~$v_y$ in this two-dimensional latent subspace to verify that the model has indeed learned a simple rotated representation of the velocity vector (Figs.\ \ref{fig:paramfit}(e), \ref{fig:paramfit}(f)).

\section{Performance Scaling with Dataset Size\label{apx:datasize}}

Due to the significant physics-informed inductive biases in our architecture, our model still achieves usable results even when trained on very small datasets. We test the dataset size dependence of our method using the 1D Kuramoto--Sivashinsky system and find that the model is still able to identify the relevant latent parameter even with a dataset of just 10 examples (Fig.\ \ref{fig:datasizetest}). The accuracy and precision of the extracted parameter and the prediction performance do begin to suffer when using such extremely small datasets, but the model is still able to provide some insight into the dynamics of the spatiotemporal system represented by the data.

\section{Alternative Boundary Conditions\label{apx:altboundary}}

The fully convolutional structure of the propagating decoder (PD) means that we are able to evaluate our model on arbitrary geometries and boundary conditions. By training on small crops and evaluating on the full size examples in the test set (Sec.\ \ref{sec:exp}), we have already shown the trained model can be directly evaluated on larger system sizes. To show direct evaluation on an alternative boundary condition, we test the refined predictive model---originally trained on the 1D Kuramoto--Sivashinsky dataset with periodic boundaries---on a new test example generated with Dirichlet hard wall boundary conditions (Fig.\ \ref{fig:KS7dirichlet}). In general, we can apply alternative boundary conditions by adjusting the padding scheme of each propagation step in the PD. For Dirichlet boundaries, this corresponds to applying anti-reflection padding at each propagation step. This preliminary test suggests that we can achieve similar prediction performance using an alternative boundary condition, which the model has never previously seen, and demonstrates the transferability of the learned convolutional kernels.

\section{Nonlinear Fiber Parameter Analysis\label{apx:fiberanalysis}}

In the three-dimensional relevant latent space $(z_1, z_2, z_3)$ extracted by the model trained on the nonlinear fiber propagation dataset (Fig.\ \ref{fig:fiber}b), we determine two independent and interpretable directions by a linear fit: $(-0.101,  0.971,  0.218)$ and $(0.477, -0.0303, -0.878)$, which correspond to the group velocity $k_1 = 1/v_g$ and second-order dispersion $k_2$, respectively (Fig.\ \ref{fig:fiber}c). The final direction $(0.882, -0.107, 0.459)$ in the latent space---orthogonal to the previous two---is a seemingly spurious relevant parameter unrelated to the parameters of the effective equation (\ref{eqn:fiber}) and represents a spurious phase velocity (Fig.\ \ref{fig:fiber}d).

For the examples in the nonlinear fiber propagation dataset, higher order dispersion terms $k_n$ are still significant. However, because these terms are correlated with $k_1$ and $k_2$ in the data, the model does not require additional latent parameters to capture their effect. Instead, the existing latent parameters also adjust the higher order dispersion terms; in other words, the latent parameters each correspond to a dispersion operator that includes higher order dispersion along with $k_1$ and $k_2$.

\bibliography{biblio}

\end{document}